# Wound Severity Classification using Deep Neural Network

D. M. Anisuzzaman[1], Yash Patel[1], Jeffrey Niezgoda[2], Sandeep Gopalakrishnan[3], and Zeyun Yu[1, 4*]

[1]Department of Computer Science, University of Wisconsin-Milwaukee, Milwaukee, WI, USA;
[2]Advancing the Zenith of Healthcare (AZH) Wound and Vascular Center, Milwaukee, WI, USA;
[3]College of Nursing, University of Wisconsin Milwaukee, Milwaukee, WI, USA;
[4]Department of Biomedical Engineering, University of Wisconsin-Milwaukee, Milwaukee, WI, USA.

[*]Corresponding authors:
Zeyun Yu, Professor and Director of Big Data Analytics and Visualization Laboratory, Department of Computer Science, University of Wisconsin-Milwaukee, Milwaukee, WI, USA. Email: yuz@uwm.edu

## Abstract

The classification of wound severity is a critical step in wound diagnosis. An effective classifier can help wound professionals categorize wound conditions more quickly and affordably, allowing them to choose the best treatment option. This study used wound photos to construct a deep neural network-based wound severity classifier that classified them into one of three classes: green, yellow, or red. The green class denotes wounds still in the early stages of healing and are most likely to recover with adequate care. Wounds in the yellow category require more attention and treatment than those in the green category. Finally, the red class denotes the most severe wounds that require prompt attention and treatment. A dataset containing different types of wound images is designed with the help of wound specialists. Nine deep learning models are used with applying the concept of transfer learning. Several stacked models are also developed by concatenating these transfer learning models. The maximum accuracy achieved on multi-class classification is 68.49%. In addition, we achieved 78.79%, 81.40%, and 77.57% accuracies on green vs. yellow, green vs. red, and yellow vs. red classifications for binary classifications.

## I. Introduction

According to a 2018 retrospective investigation [1], more than 8 million people had wounds, and the Medicare cost of wound treatments varied from $28.1 billion to $96.8 billion. This notable figure can give us an indication of the number of people who are affected by wounds and how they are cared for and managed. Diabetic foot ulcers (DFU), venous leg ulcers (VLU), pressure ulcers (PU), and surgical wounds (SW) are the most prevalent forms of wounds/ulcers. The DFU affects around 34% of patients with diabetes during their lifetime, and more than 50% of diabetic foot ulcers become infected [2]. Active VLU affects 0.15 percent to 0.3 percent of the world's population [3]. A pressure ulcer is another serious lesion that affects 2.5 million individuals each year [4]. Approximately 4.5 percent of adults have surgery that results in a surgical wound each year [5].

According to the data above, wounds have resulted in a significant financial burden for patients and may even be life-threatening. In addition, many wound patients lack access to specialized wound care and the latest recommendations due to a scarcity of well-trained wound experts in primary and rural healthcare settings. Therefore, remote telemedicine system advancements can considerably assist patients in remote regions, particularly in rural areas, by providing improved diagnostic guidance, which is especially important in pandemics like COVID-19 [6]. Furthermore, with the growing usage of artificial intelligence (AI) technology and portable devices such as smartphones, it is now more essential to build remote and intelligent wound care diagnostic and prognostic systems. In wound care, an intelligent system can help in various ways, including enhanced accuracy, decreased effort and budgetary burden, standardized diagnosis and management, and improved patient care quality [7].

Wound classification, which includes wound severity detection, differentiating between different types of wounds (DFU, VLU, PU, SW, etc.) and wound conditions (infection vs. non-infection, ischemia vs. non-ischemic, etc.); wound tissue classification, burn depth classification, and so on, is an important part of this wound care system. Physicians must first determine the type of wound to provide appropriate medication and treatment instructions. Wound specialists manually classified wounds before AI growth. As a result, AI can save time and money while also making better predictions than humans in some circumstances [8]. In comparison to the early generations of rule-based AI, which relied heavily on an expert's knowledge, AI algorithms have evolved in recent years into so-called data-driven systems that do not require human or expert participation [9].

Wound severity classification is an essential part of the wound diagnosis process as this study can help physicians make quick and proper decisions on treatment plan making. There are three classes named green, yellow, and red. Green represents the wounds in the primary stage and is most likely to heal with proper treatment. The yellow class contains those wounds that need more attention and treatment than those in the green category. Finally, the red class contains the most severe wounds that require immediate action and quick and proper treatment. The characteristics upon which wound severity can be determined are shown in Figure 1. Color, peri-wound region, size, and depth of wounds are the indicating factors that can differentiate among the green, yellow, and red classes.

This research is the first study that measures and classifies wound severity from wound images using deep learning models to the best of our knowledge. The related works, dataset collection, dataset preparation, model training process, and experiments done to perform this classification among red, yellow, and green images are discussed in the following sections.

| Characteristic | Red | Yellow | Green |
|---|---|---|---|
| Color | | | |
| Red 100% | | | X |
| Yellow-Grey <50% | | X | |
| Yellow-Grey 50-100% | X | | |
| Black-Brown | X | | |
| Peri-wound | | | |
| Normal | | | X |
| Callus | | X | |
| Red <1cm | | X | |
| Red >1cm | X | | |
| Maceration | | X | |
| Maceration and Breakdown | X | | |
| Size | | | |
| 2cm or less | | | X |
| 2 – 5 cm | | X | |
| >5 cm | X | | |
| Depth | | | |
| Minimal-None | | | X |
| 1cm or less | | X | |
| >1cm | X | | |

**Figure 1:** Photo characteristics of Red-Yellow-Green stratification.

## II. Related Works

Acute and chronic wound classification includes wound type classification, tissue classification, severity classification, etc. Wound type classification considers different types of wounds and non-wounds (normal skin, background, etc.). Normal skin versus PU and DFU versus PU are binary wound type classification examples. DFU versus PU versus VLU is an example of multi-class wound type classification. Granulation versus slough versus necrosis is an example of wound tissue classification.

Several wound type classification works used machine learning and deep learning-based methods. Goyal et al. [10] used traditional machine learning, deep learning, and ensemble CNN models for binary classification of ischemia versus non-ischemia and infection versus non-infection on DFU images. Abubakar et al. [11] proposed a machine learning approach to differentiating burn wounds and pressure ulcers. A novel CNN architecture named DFUNet was developed by Goyal et al. [12] for binary classification of healthy skin and DFU skin. Shenoy et al. [13] proposed a CNN-based method for binary classification of wound images. A binary patch classification of normal skin versus abnormal skin (DFU) is performed by Alzubaidi et al. [14] with a novel deep convolutional neural network named DFU_QUTNet. A CNN-based method is proposed by Aguirre et al. [15] for VLU versus non-VLU classification from ulcer images. Rostami et al. [16] proposed an end-to-end ensemble DCNN-based classifier to classify the entire wound images into multiple classes, including surgical, diabetic, and venous ulcers. Sarp et al. [17] classified chronic wounds into four classes (diabetic, lymphovascular, pressure injury, and surgical) by using an explainable artificial intelligence (XIA) approach to provide transparency on the neural network. Anisuzzaman et al. [18] proposed a multimodality-based deep learning network to classify DFU, PU, VLU, and SW using wound images and corresponding locations.

Machine learning and deep learning-based approaches were also employed in various wound tissue classification studies. Wannous et al. [19] developed a multi-view strategy for tissue classification (granulation, slough, and necrosis) based on an SVM classifier. Veredas et al. [20] classified five types of tissue (necrotic, slough, granulation, healing, and skin) in wound regions using a hybrid approach consisting of neural networks and Bayesian classifiers. Wannous et al. [21] classified three types of wound tissue (granulation, slough, and necrosis) using an SVM region classifier, with color and texture as input. Mukherjee et al. [22] classified three types of wound tissue (granulation, necrotic, and slough) using Bayesian and SVM classifiers. Wannous et al. [23] classified four tissue types (granulation, slough, necrosis, and healthy skin) using color and texture features as input to the SVM region classifier. Zahia et al. [24] proposed a method for tissue analysis of pressure wound images using deep CNNs. Rajathi et al. [25] proposed a CNN-based method for tissue classification of varicose ulcer wound images. Nejati et al. [26] proposed a deep learning-based method to analyze chronic wound tissue and classify it into one of the seven classes: necrotic, healthy granulation, slough, infected, unhealthy granulation, hyper granulation, epithelialization. Rania et al. [27] performed wound tissue classification into three classes: necrosis, granulation, and slough, by using a superpixel-wise FCN approach.

Though there are many works on wound type and wound tissue classifications, to the best of our knowledge, there is no existing work that uses wound images to learn features to classify them according to their severity. The most related work that can be slightly compared to our work is performed by Nguyen et al. [28]. They explored different machine learning classifiers to provide wound care decisions (target class). The target classes are: 1) Continue current treatment, 2) Request a non-urgent change in treatment from a wound specialist, and 3) Refer the patient to a wound specialist. They used visual wound features and unstructured text from wound experts as the input of their machine learning models. Using only visual features, they achieved an accuracy of 76% with the Support Vector Machine (SVM) classifier. Using both visual and textual wound features, they improved accuracy to 81% with the Gradient Boosted Machine (XGBoost).

The main difference between [28] and our work is that we did not use any textual features, and the visual features authors described were not learned from the images using machine learning methods. The visual features are collected by a visiting nurse or a nursing home nurse. The nurse would capture a photo of the wound with their developed smartphone app and be prompted to enter some wound information such as wound location, appearance, and other clinical characterizations that a nurse could easily assess visually. Where in our work, all the features were learned through convolution to do the classification among green, yellow, and red classes. This makes our developed system free from any wound expert input, which is the primary key to automation and a remote wound care system.

## III. Methodology

### 3.1 Dataset Collection

The dataset was collected from the AZH Wound and Vascular Center, Milwaukee, WI, USA. This dataset contains a total of 420 wound images of different types of wounds, including diabetic, pressure, and venous ulcers. Among these images, 100 are in the green class, 175 are in the yellow class, and 145 belong to the red class. An expert wound specialist from the AZH center classified the images into these classes. All the images were captured with iPad and DSLR cameras. No specific environmental or illumination condition was applied during image capturing. These images were further processed and used as training, validation, and test data. Some sample images of each class are shown in Figure 2.

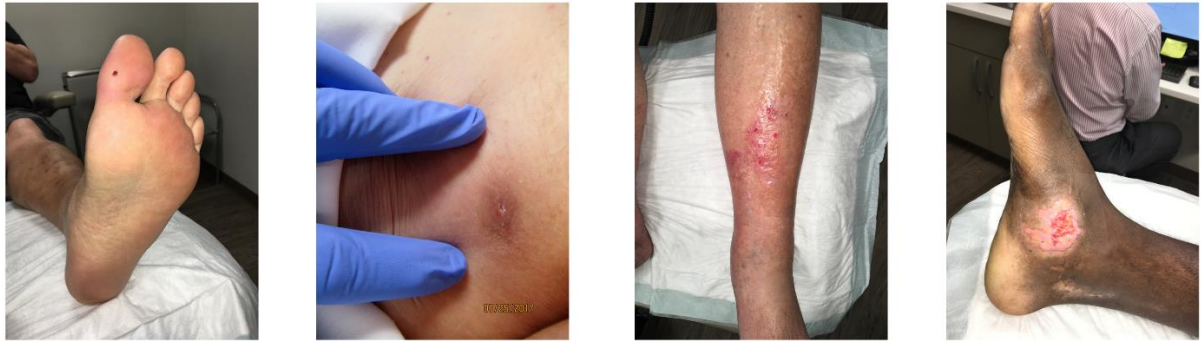

(a) Green

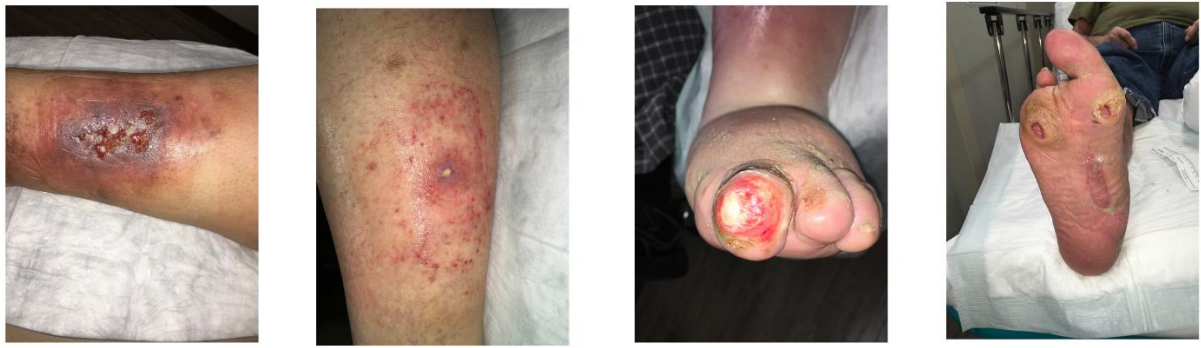

(b) Yellow

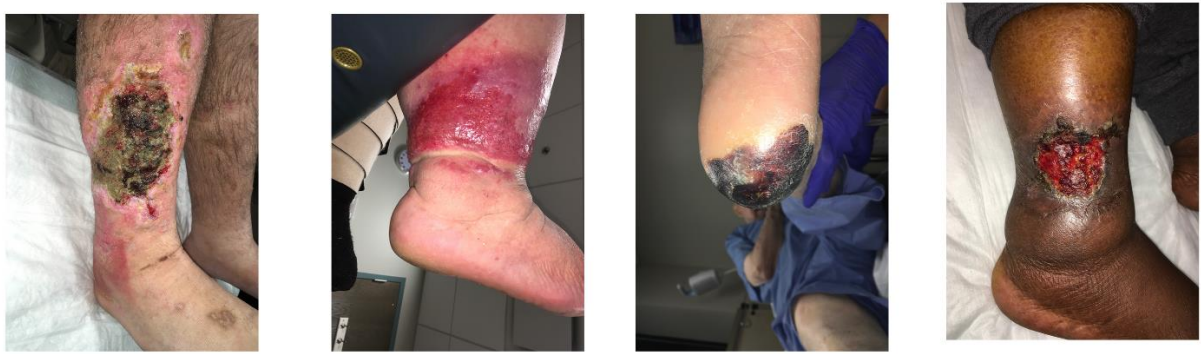

(c) Red

**Figure 2:** Wound severity database sample images. (a), (b) and (c) rows represent the examples of green, yellow, and red classes, respectively.

## 3.2 Dataset Preparation

First, all the images were passed through our developed wound localizer [29] to select the region of interest (ROIs). The developed wound localizer uses the YOLOv3 model to predict the ROIs with a confidence score (mAP value) greater than 97%. The developed localizer automatically cropped the ROIs and passed them to the severity classifier. We have 723 ROIs from the original images as a single image contains multiple wounds. 80% of the ROIs were used for training and validation, and 20% for testing. The splitting was done carefully so that there should be no overlap between them.

From Figure 1, we can see that the peri-wound area has a significant role in the characteristics of the three classes (green, red, and yellow). The peri-wound region is defined as the area of skin that extends beyond the wound edge to a certain amount. To learn the features from the peri-wound area, we applied three zoom-outs on the original ROIs (training and validation set only) produced by our developed YOLOv3 model [29]. The used terms for this experiment are Z0: zoom out zero or original ROIs, Z1: zoom out one with 50 pixels padding on the original ROIs, Z2: zoom out two with 100 pixels padding on the original ROIs, and Z3: zoom out three with 150 pixels padding on the original ROIs. Then, the training set of each Z0 to Z3 was augmented by using horizontal flip, vertical flip, and three rotations (25, 45, and 90 degrees). The dataset preparation process for wound severity classification is shown in Figure 3. Table 1 shows the dataset splitting and augmentation statistics for wound severity classification.

**Table 1:** Database summary for wound severity classification.

| Class | ROIs | | | Augmented ROIs | | |
|---|---|---|---|---|---|---|
| | Train + Validation | Test | Total | Train + Validation | Test | Total |
| Green | 154 | 39 | **193** | 924 | 39 | **963** |
| Red | 237 | 60 | **297** | 1116 | 60 | **1176** |
| Yellow | 186 | 47 | **233** | 1422 | 47 | **1469** |
| **Total** | **577** | **146** | **723** | **3462** | **146** | **3608** |

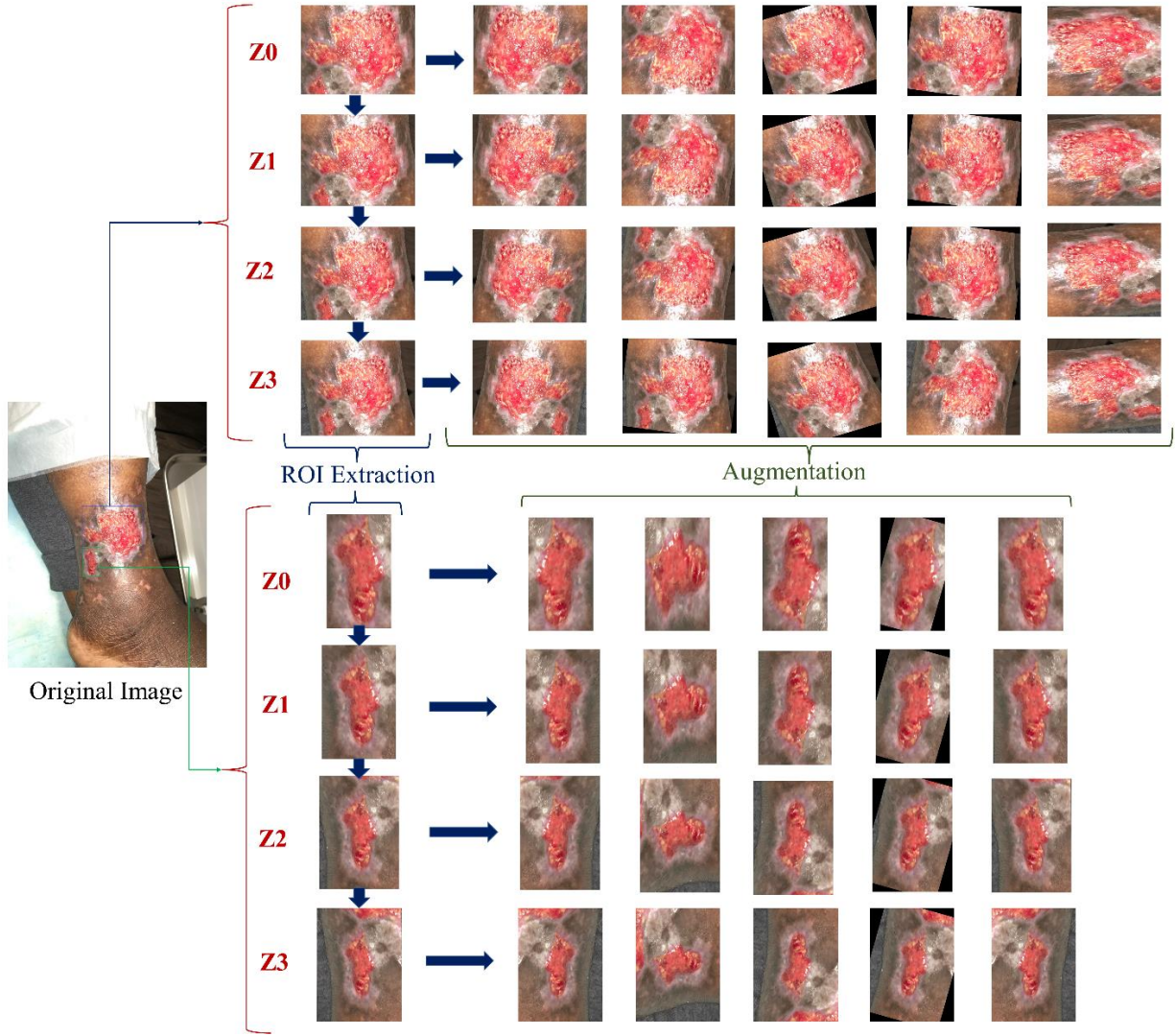

**Figure 3:** Dataset processing steps for wound severity classification.

### 3.3 Models

Transfer learning and stacked deep learning models were used to evaluate this dataset. Transfer learning means taking advantage of features learned on one problem and using them in another similar situation. This method is proper when the dataset in hand is small in number to train a full-scale model from scratch, and the memory power is limited to train a vast deep learning model. The most used workflow of transfer learning is: 1) take a previously trained model's layers, 2) freeze the layers, 3) add some new, trainable layers on top of the frozen layers, which will learn to turn the old features into predictions on a new dataset, and 4) train the new layers on the new dataset [30]. There are 26 deep learning models in Keras Applications [31], among which we choose nine classification models: VGG16 [32], VGG19 [32], InceptionV3 [33], NasNetLarge [34], ResNet50 [35], DenseNet201 [36], Xception [37], MobileNetV2 [38], and InceptionResNetV2 [39]. All nine models were pre-trained on ImageNet [40], a huge dataset including more than 14 million general images of 1000 classes.

Stacked models are a mixture of these nine individual models. These stacked models include two different models trained on the same dataset. The stacked models were used in the hope that different models may learn some new features that the other model missed and vice versa. Six stacked models were used, and the individual models were picked based on their standalone results. After images went through each network, their outputs were concatenated, and four dense layers were added on top of it to learn about these mixed features, followed by a SoftMax (output) layer.

Finally, nine stacked models were trained, with each having four individual models. These models are different from the previous stacked models because they do not pass the same image through each individual network. Here, four separate networks were trained on the same image's zoom-out channels (Z0 to Z3). These nine stacked models are called multi-zoom learning networks. After images went through every four individual networks, their outputs were concatenated, and five dense layers were added on top of it to learn about these mixed features, followed by a SoftMax (output) layer.

## IV. Experiment and Results

### 4.1 Experimental Setup

One multi-class classification among green, yellow, and red classes was performed. This classification was performed by using the original ROIs (Z0) and their zoom-out images (Z1, Z2, and Z3). A multi-zoom classification was also conducted using all the zoom-out (Z0 to Z3) channels of a single ROI. We also performed three binary classifications: green vs. yellow, green vs. red, and yellow vs. red.

All models were written in Python programming language by using Tensorflow.Keras deep learning framework and trained on an Nvidia GeForce RTX 2080Ti GPU platform. With a learning rate of 0.001 and an Adam optimizer, all models were trained for 250 epochs. Two callbacks were employed with the best validation accuracy and the optimal combination of validation and training accuracy saving. The "*sparse_categorical_crossentropy*" and "*binary_crossentropy*" loss functions were used for multi-class and binary classification, respectively.

We use accuracy as the performance metric to investigate the classification performance. The confusion matrix of the best model for classifying Z0 images is also provided. Accuracy is the ratio of correctly predicted data to the total amount of data. Equation 1 shows the formula for accuracy. In this equation, TP, TN, FP, and FN, represent True Positive, True Negative, False Positive, and False Negative measures. More details about this equation can be found in [41].

$$Accuracy = \frac{TP + TN}{TP + FP + FN + TN} \quad (1)$$

## 4.2 Results

Nine transfer learning models and six stacked models were used to classify the original ROIs (Z0). The result of this experiment is shown in Table 2.

**Table 2:** Wound severity classification performance on original ROIs (Z0 images).

| Model Type | Model | Accuracy |
|---|---|---|
| **Transfer Learning** | VGG16 | 60.27% |
| | VGG19 | **68.49%** |
| | InceptionV3 | 52.05% |
| | NasNetLarge | 56.85% |
| | ResNet50 | 41.10% |
| | DenseNet201 | 41.10% |
| | Xception | 56.16% |
| | MobileNetV2 | 49.32% |
| | InceptionResNetV2 | 47.26% |
| **Stacked Models** | M1: VGG19+NasNetLarge | 67.81% |
| | M2: NasNetLarge + Xception | 60.96% |
| | M3: VGG19+InceptionV3 | 64.38 % |
| | M4: VGG16+NasNetLarge | 65.75% |
| | M5: InceptionV3 + Xception | 56.16% |
| | M6: VGG16+InceptionV3 | 63.70% |

Training all these nine individual and six stacked models is expensive in terms of both time and cost. So, we selected five individual models and three stacked models based on their results achieved on the original ROI (Z0) classification to perform classification on zoom-out images (Z1 to Z3). The accuracy of classifying these Z1, Z2, and Z3 images are shown in Table 3.

**Table 3:** Wound severity classification performance on zoom-out (Z1, Z2, and Z3) images.

| Model Type | Model | Accuracy | | |
|---|---|---|---|---|
| | | Z1 | Z2 | Z3 |
| Transfer Learning | VGG16 | 62.33% | 60.96% | 55.48% |
| | VGG19 | 66.44% | 58.22% | 61.64% |
| | InceptionV3 | 56.16% | 56.85% | 53.42% |
| | NasNetLarge | 56.85% | 61.64% | 57.53% |
| | Xception | 56.85% | 51.37% | **63.01%** |
| Stacked Models | **M1:** VGG19+NasNetLarge | 67.12% | 58.90% | 58.22% |
| | **M3:** VGG19+InceptionV3 | 63.70% | 56.85% | 58.90% |
| | **M4:** VGG16+NasNetLarge | **68.49%** | **63.01%** | 55.48% |

**Table 4:** Multi-zoom learning results for wound severity classification.

| | Stacked Model | Accuracy |
|---|---|---|
| **M1** | **Z0:** VGG19; **Z1:** InceptionV3; **Z2:** NasNet large; **Z3:** Res Net 50 | 58.90% |
| **M2** | **Z0:** Res Net 50; **Z1:** InceptionV3; **Z2:** NasNet large; **Z3:** VGG19 | 58.22% |
| **M3** | **Z0:** Res Net 50; **Z1:** VGG16; **Z2:** NasNet large; **Z3:** InceptionV3 | 56.85% |
| **M4** | **Z0:** Res Net 50; **Z1:** InceptionV3; **Z2:** NasNet large; **Z3:** Xception | 52.74% |
| **M5** | **Z0:** VGG19; **Z1:** InceptionV3; **Z2:** NasNet large; **Z3:** MobileNetV2 | **63.70%** |
| **M6** | **Z0:** VGG19; **Z1:** InceptionResNetV2; **Z2:** NasNet large; **Z3:** MobileNetV2 | 62.33% |
| **M7** | **Z0:** DenseNet 201; **Z1:** InceptionResNetV2; **Z2:** NasNet large; **Z3:** MobileNetV2 | 54.11% |
| **M8** | **Z0:** Xception; **Z1:** InceptionResNetV2; **Z2:** DenseNet 201; **Z3:** MobileNetV2 | 50.68% |
| **M9** | **Z0:** VGG19; **Z1:** InceptionResNetV2; **Z2:** Res Net 50; **Z3:** MobileNetV2 | 60.27% |

Nine stacked models, each having four individual models, were trained for multi-zoom learning, where four individual networks were trained on four zoom-out versions (Z0 to Z3) of the same image. The individual models include all nine transfer learning models. We cannot apply the same individual models twice or more because, during concatenation, each layer's name must be unique. Table 4 shows the result of this multi-zoom learning. Here, the model's name followed by the zoom-out channel (Z0 to Z3) was trained on the images of that specific zoom-out channel.

Finally, we performed three binary classifications on this dataset. The classification performed are green versus yellow, green versus red, and yellow versus red. We used five individual and three stacked models based on their original ROI (Z0) multi-class classification results to perform these three binary classifications. Table 5 shows the results of binary classifications on the wound severity database. The binary classifications were performed on the original ROIs (Z0 images).

**Table 5:** Binary classification results on wound severity dataset (Original ROIs).

| Model Type | Model | Accuracy | | |
| --- | --- | --- | --- | --- |
| | | Green Vs. Yellow | Green Vs. Red | Yellow Vs. Red |
| Transfer Learning | VGG16 | 71.72% | 77.91% | 73.83% |
| | VGG19 | 76.77% | 76.74% | 75.70% |
| | InceptionV3 | 63.64% | 80.23% | 64.49% |
| | NasNetLarge | 64.65% | **81.40%** | 72.90% |
| | Xception | 70.71% | 65.12% | 68.22% |
| Stacked Models | **M1:** VGG19+NasNetLarge | 65.66% | 74.42% | 72.90% |
| | **M3:** VGG19+InceptionV3 | **78.79%** | 77.91% | 68.22% |
| | **M4:** VGG16+NasNetLarge | 63.64% | 77.91% | **77.57%** |

# V.  Discussion

From Table 2, we can see that the VGG19 model achieved the highest accuracy (68.49%). The second highest accuracy (60.27%) achieved by any individual model was VGG16, the previous version of the VGG19 model. From the stacked models, the highest accuracy (67.81%) was acquired by VGG19 and NasNetLarge combination. Though we expect more accuracy from these stacked models, they cannot outperform the VGG19 model. But in general, stacked models did well than individual models. For example, VGG16 and NasNetlarge have individual accuracy of 60.27% and 56.85%; but their stacked model achieved an accuracy of 65.75%.

From Table 3, we can see that using zoom-out does not improve the classification performance. Zoom-out one (Z1) with 50 pixels padding images produced the same accuracy as Z0. In the case of Z2 and Z3, we can see a reduction in accuracy value. Though we are using zoom-out images to capture features from the peri-wound regions, increasing the padding size (100 pixels for Z2 and 150 pixels for Z3) of ROIs may capture some unnecessary information like non-skin regions (including bed sheets, the hand of physicians, wall, floor, etc.), which in turns decreased the accuracy. For Z1 and Z2, the highest accuracy was obtained by the M4 stacked network, which is a combination of VGG16 and NasNetLarge networks. The Xception network achieved the highest accuracy for Z3 images.

From Table 4, multi-zoom learning is not helping in the wound severity classification. The highest accuracy was achieved by the M5 model, where Z0 images were passed through the VGG19 network, Z1 images were passed through the InceptionV3 network, Z2 images were passed through the NasNetLarge network, and Z3 images were passed through the MobileNetV2 network. The highest accuracy of multi-zoom learning is 63.70%, almost 5% lower than the accuracy achieved with the original ROIs. As discussed in the above paragraph, the unnecessary information from Z2 and Z3 images may have reduced the classification performance.

From Table 5, the highest accuracy achieved for three binary classifications are 78.79%, 81.40%, and 77.57% for green vs. yellow, green vs. red, and yellow vs. red, respectively. The M3 and M4 stacked models achieved the highest accuracies for green vs. yellow and yellow vs. red. For green vs. red, the highest accuracy was acquired by the NasNetlarge network. This result shows that without the middle class (yellow), the green and red images are easier to distinguish.

The confusion matrix for the best model (VGG19) of multi-class wound severity image classification on original ROIs (Z0) is shown in Figure 4. The most challenging task was distinguishing between green vs. yellow images and yellow vs. red images from this figure. For this reason, as the middle class, yellow has the highest class-wise recall and lowest class-wise precision. From Figure 4, we can see that only 2 green images were misclassified as red, and only one red image was misclassified as green. Some examples of misclassification by the best model of multi-class wound severity image classification on original ROIs are shown in Figure 5.

|  | | Gold Label | | | |
|---|---|---|---|---|---|
|  | | Green | Yellow | Red | Precision |
| Prediction | Green | 25 | 6 | 1 | 78.1% |
| | Yellow | 12 | 47 | 18 | 61.0% |
| | Red | 2 | 7 | 28 | 75.7% |
| | Recall | 64.1% | 78.3% | 59.6% | **68.5%** |

**Figure 4:** Confusion matrix for the best model (VGG19) of multi-class wound severity image classification on the original ROIs.

From Figure 5, we can see that it is hard to distinguish between the three classes (green, yellow, and red) from a general human perspective. For example, the second ROI in the top row of Figure 5 was annotated as green, but it generally looks like red. Also, the third ROI in the top row of Figure 5 was annotated as yellow, but it looks like green. So, from the physician's point of view (who annotated the wound severity dataset), there may be some other features than the standard features (i.e., color, texture, edge, etc.) that are not captured by the convolution and deep neural networks. This statement can also be supported by Figure 1. We are now considering only the 2D image features learned through convolutions, but from Figure 1, we can see that there are other contributing factors in this classification task (e.g., size, depth, maceration, breakdown, callus tissue, etc.) that must be incorporated to improve the classification performance. To incorporate these contributing factors (e.g., size, depth, callus tissue, etc.), we may need the output from other wound automation modalities like wound segmentation, wound tissue classification, etc. Finally, we have only 420 wound images, insufficient to develop a sophisticated deep learning model like a wound severity classifier. Also, the collected images on the dataset have lighting issues that hamper the classification performance. The images on the dataset were also not taken from a fixed distance, making the work of wound size calculation much harder. Addressing all these issues and developing an improved and rich dataset in the future will improve the wound severity classification performance by a good margin.

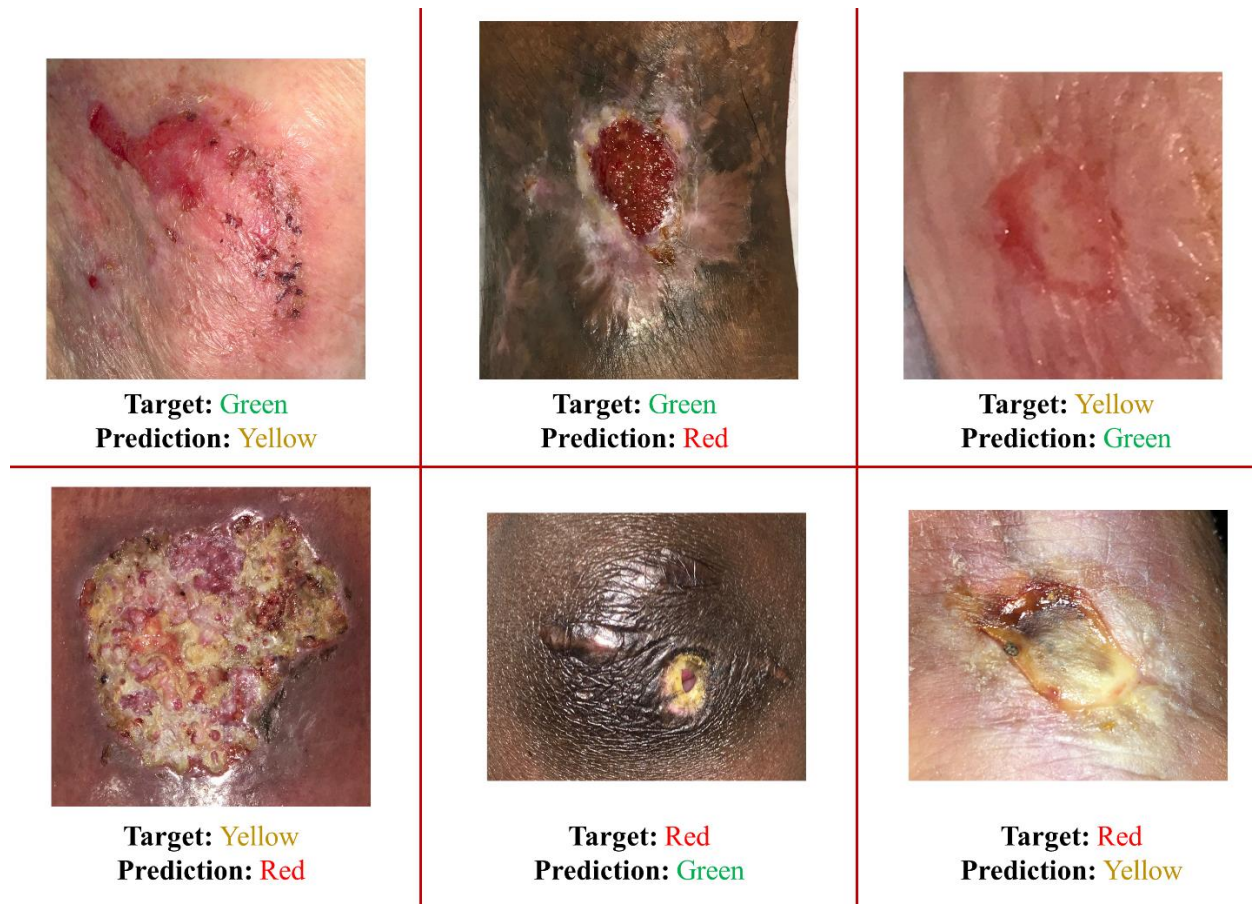

**Figure 5:** Examples of wound severity image misclassifications by the best model (VGG19) for multi-class classification on the original ROIs (Z0).

As discussed in section 2 (related works), we do not have any existing work (to the best of our knowledge) to compare with our current research. The work performed by Nguyen et al. [28] is not reasonably comparable to our method, as they had used hand-crafted features (by a wound expert: nurse) for their machine learning models. The features used in our work are extracted automatically by deep learning models. This makes our work completely automated and ready to use without any expert's input, which is extremely useful for remote assessment of wounds. Also, the classes used on [28] (continue current treatment, request a non-urgent change in treatment from a wound specialist, and refer patient to a wound specialist) may represent different characteristics from the classes (green, yellow, and red) we used in this research.

# VI. Conclusion

In this research, several deep learning networks are applied to a novel dataset collected from the AZH wound center for wound severity classification. This work can help wound professionals in treatment plan making. One multi-class and three binary classifications are performed on this developed dataset to differentiate among wound severity levels. We achieved an accuracy of 68.49% for multi-class classification and 77.57% to 81.40% accuracy for binary classifications. The reason behind this poor performance is the relevant factors (size, depth, etc.) which cannot be measured from a 2D image through just convolution. Also, the scarcity of data plays an essential role in model performance. In the future, these relevant factors should be integrated into the model, and the dataset should be polished and increased to improve wound severity classification performance.

# Reference

[1]  C. K. Sen, "Human Wound and Its Burden: Updated 2020 Compendium of Estimates," *Adv. Wound Care*, vol. 10, no. 5, pp. 281–292, Mar. 2021, doi: 10.1089/WOUND.2021.0026.

[2]  "Diabetic Foot: Facts & Figures." https://diabeticfootonline.com/diabetic-foot-facts-and-figures/ (accessed Jun. 02, 2021).

[3]  E. A. Nelson and U. Adderley, "Venous leg ulcers," *BMJ Clin. Evid.*, vol. 2016, no. March 2014, pp. 1–36, 2016.

[4]  "Preventing Pressure Ulcers in Hospitals." https://www.ahrq.gov/patient-safety/settings/hospital/resource/pressureulcer/tool/pu1.html (accessed Jun. 04, 2021).

[5]  B. M. Gillespie *et al.*, "Setting the surgical wound care agenda across two healthcare districts: A priority setting approach," *Collegian*, vol. 27, no. 5, pp. 529–534, 2020, doi: 10.1016/j.colegn.2020.02.011.

[6]  F. L. Bowling *et al.*, "Remote assessment of diabetic foot ulcers using a novel wound imaging system," *Wound Repair Regen.*, vol. 19, no. 1, pp. 25–30, Jan. 2011, doi: 10.1111/J.1524-475X.2010.00645.X.

[7]  E. Pasero and C. Castagneri, "Application of an automatic ulcer segmentation algorithm," in *RTSI 2017 - IEEE 3rd International Forum on Research and Technologies for Society and Industry, Conference Proceedings*, Oct. 2017, pp. 1–4, doi: 10.1109/RTSI.2017.8065954.

[8]  D. M. Anisuzzaman, C. Wang, B. Rostami, S. Gopalakrishnan, J. Niezgoda, and Z. Yu, "Image-Based Artificial Intelligence in Wound Assessment: A Systematic Review," *Adv. Wound Care*, Dec. 2021, doi: 10.1089/WOUND.2021.0091.

[9]  K. H. Yu, A. L. Beam, and I. S. Kohane, "Artificial intelligence in healthcare," *Nat. Biomed. Eng.*, vol. 2, no. 10, pp. 719–731, 2018, doi: 10.1038/s41551-018-0305-z.

[10] M. Goyal, N. D. Reeves, S. Rajbhandari, N. Ahmad, C. Wang, and M. H. Yap, "Recognition of ischaemia and infection in diabetic foot ulcers: Dataset and techniques," *Comput. Biol. Med.*, vol.

117, p. 103616, Feb. 2020, doi: 10.1016/J.COMPBIOMED.2020.103616.

[11] A. Abubakar, H. Ugail, and A. M. Bukar, "Can Machine Learning Be Used to Discriminate Between Burns and Pressure Ulcer?," *Adv. Intell. Syst. Comput.*, vol. 1038, pp. 870–880, Sep. 2019, doi: 10.1007/978-3-030-29513-4_64.

[12] M. Goyal, N. D. Reeves, A. K. Davison, S. Rajbhandari, J. Spragg, and M. H. Yap, "DFUNet: Convolutional Neural Networks for Diabetic Foot Ulcer Classification," *IEEE Trans. Emerg. Top. Comput. Intell.*, vol. 4, no. 5, pp. 728–739, Sep. 2018, doi: 10.1109/TETCI.2018.2866254.

[13] V. N. Shenoy, E. Foster, L. Aalami, B. Majeed, and O. Aalami, "Deepwound: Automated Postoperative Wound Assessment and Surgical Site Surveillance through Convolutional Neural Networks," in *Proceedings - 2018 IEEE International Conference on Bioinformatics and Biomedicine, BIBM 2018*, Jan. 2019, pp. 1017–1021, doi: 10.1109/BIBM.2018.8621130.

[14] L. Alzubaidi, M. A. Fadhel, S. R. Oleiwi, O. Al-Shamma, and J. Zhang, "DFU_QUTNet: diabetic foot ulcer classification using novel deep convolutional neural network," *Multimed. Tools Appl. 2019 7921*, vol. 79, no. 21, pp. 15655–15677, Jun. 2019, doi: 10.1007/S11042-019-07820-W.

[15] C. A. NILSSON and M. VELIC, "Classification of Ulcer Images Using Convolutional Neural Networks," 2018.

[16] B. Rostami, D. M. Anisuzzaman, C. Wang, S. Gopalakrishnan, J. Niezgoda, and Z. Yu, "Multiclass wound image classification using an ensemble deep CNN-based classifier," *Comput. Biol. Med.*, vol. 134, p. 104536, Jul. 2021, doi: 10.1016/J.COMPBIOMED.2021.104536.

[17] S. Sarp, M. Kuzlu, E. Wilson, U. Cali, and O. Guler, "A Highly Transparent and Explainable Artificial Intelligence Tool for Chronic Wound Classification: XAI-CWC," no. January, pp. 1–13, 2021, doi: 10.20944/preprints202101.0346.v1.

[18] D. M. Anisuzzaman, Y. Patel, B. Rostami, J. Niezgoda, S. Gopalakrishnan, and Z. Yu, "Multi-modal Wound Classification using Wound Image and Location by Deep Neural Network," Sep. 2021, doi: 10.48550/arxiv.2109.06969.

[19] H. Wannous, Y. Lucas, and S. Treuillet, "Enhanced assessment of the wound-healing process by accurate multiview tissue classification," *IEEE Trans. Med. Imaging*, vol. 30, no. 2, pp. 315–326, Feb. 2011, doi: 10.1109/TMI.2010.2077739.

[20] F. Veredas, H. Mesa, and L. Morente, "Binary tissue classification on wound images with neural networks and bayesian classifiers," *IEEE Trans. Med. Imaging*, vol. 29, no. 2, pp. 410–427, Feb. 2010, doi: 10.1109/TMI.2009.2033595.

[21] H. Wannous, S. Treuillet, and Y. Lucas, "Supervised tissue classification from color images for a complete wound assessment tool," *Annu. Int. Conf. IEEE Eng. Med. Biol. - Proc.*, pp. 6031–6034, 2007, doi: 10.1109/IEMBS.2007.4353723.

[22] R. Mukherjee, D. D. Manohar, D. K. Das, A. Achar, A. Mitra, and C. Chakraborty, "Automated tissue classification framework for reproducible chronic wound assessment," *Biomed Res. Int.*, vol. 2014, 2014, doi: 10.1155/2014/851582.

[23] H. Wannous, Y. Lucas, and S. Treuillet, "Efficient SVM classifier based on color and texture region features for wound tissue images," *Med. Imaging 2008 Comput. Diagnosis*, vol. 6915, p. 69152T, Mar. 2008, doi: 10.1117/12.770339.

[24] S. Zahia, D. Sierra-Sosa, B. Garcia-Zapirain, and A. Elmaghraby, "Tissue classification and segmentation of pressure injuries using convolutional neural networks," *Comput. Methods*

*Programs Biomed.*, vol. 159, pp. 51–58, Jun. 2018, doi: 10.1016/J.CMPB.2018.02.018.

[25] V. Rajathi, R. R. Bhavani, and G. W. Jiji, "Varicose ulcer(C6) wound image tissue classification using multidimensional convolutional neural networks," *Imaging Sci. J.*, vol. 67, no. 7, pp. 374–384, Oct. 2019, doi: 10.1080/13682199.2019.1663083.

[26] H. Nejati *et al.*, "Fine-Grained Wound Tissue Analysis Using Deep Neural Network," *ICASSP, IEEE Int. Conf. Acoust. Speech Signal Process. - Proc.*, vol. 2018-April, pp. 1010–1014, Sep. 2018, doi: 10.1109/ICASSP.2018.8461927.

[27] R. Niri, H. Douzi, Y. Lucas, and S. Treuillet, "A Superpixel-Wise Fully Convolutional Neural Network Approach for Diabetic Foot Ulcer Tissue Classification," in *In Pattern Recognition. ICPR International Workshops and Challenges: Virtual Event*, 2021, pp. 308–320.

[28] H. Nguyen *et al.*, "Machine learning models for synthesizing actionable care decisions on lower extremity wounds," *Smart Heal.*, vol. 18, p. 100139, Nov. 2020, doi: 10.1016/J.SMHL.2020.100139.

[29] D. M. Anisuzzaman, Y. Patel, J. Niezgoda, S. Gopalakrishnan, and Z. Yu, "A Mobile App for Wound Localization using Deep Learning," 2020. doi: arXiv preprint arXiv:2009.07133.

[30] F. Chollet, "Transfer learning & fine-tuning," *Keras*. https://keras.io/guides/transfer_learning/ (accessed Jul. 02, 2021).

[31] "Keras Applications." https://keras.io/api/applications/ (accessed Jul. 16, 2021).

[32] K. Simonyan and A. Zisserman, "Very Deep Convolutional Networks for Large-Scale Image Recognition," *3rd Int. Conf. Learn. Represent. ICLR 2015 - Conf. Track Proc.*, Sep. 2014, Accessed: Jul. 16, 2021. [Online]. Available: https://arxiv.org/abs/1409.1556v6.

[33] C. Szegedy *et al.*, "Going Deeper with Convolutions," in *IEEE conference on computer vision and pattern recognition*, 2015, pp. 1–9.

[34] B. Zoph, V. Vasudevan, J. Shlens, and Q. V. Le, "Learning Transferable Architectures for Scalable Image Recognition," in *IEEE conference on computer vision and pattern recognition*, 2018, pp. 8697–8710.

[35] K. He, X. Zhang, S. Ren, and J. Sun, "Deep Residual Learning for Image Recognition," in *IEEE Conference on Computer Vision and Pattern Recognition (CVPR)*, 2016, pp. 770–778, Accessed: Jul. 16, 2021. [Online]. Available: http://image-net.org/challenges/LSVRC/2015/.

[36] Gao Huang, Zhuang Liu, Laurens van der Maaten, and Kilian Q. Weinberger, "Densely Connected Convolutional Networks," in *Proceedings of the IEEE Conference on Computer Vision and Pattern Recognition (CVPR)*, 2017, pp. 4700–4708, Accessed: Mar. 29, 2022. [Online]. Available: https://openaccess.thecvf.com/content_cvpr_2017/html/Huang_Densely_Connected_Convolutional_CVPR_2017_paper.html.

[37] Francois Chollet, "Xception: Deep Learning With Depthwise Separable Convolutions," in *Proceedings of the IEEE Conference on Computer Vision and Pattern Recognition (CVPR)*, 2017, pp. 1251–1258, Accessed: Mar. 29, 2022. [Online]. Available: https://openaccess.thecvf.com/content_cvpr_2017/html/Chollet_Xception_Deep_Learning_CVPR_2017_paper.html.

[38] M. Sandler, A. Howard, M. Zhu, A. Zhmoginov, and L.-C. Chen, "MobileNetV2: Inverted Residuals and Linear Bottlenecks," 2018.

[39] C. Szegedy, S. Ioffe, V. Vanhoucke, and A. A. Alemi, "Inception-v4, Inception-ResNet and the Impact of Residual Connections on Learning," Accessed: Mar. 29, 2022. [Online]. Available: www.aaai.org.

[40] "ImageNet." https://image-net.org/update-mar-11-2021.php (accessed Apr. 04, 2022).

[41] "Accuracy, Precision, Recall & F1 Score: Interpretation of Performance Measures - Exsilio Blog," *Exsilio Solutions*. https://blog.exsilio.com/all/accuracy-precision-recall-f1-score-interpretation-of-performance-measures/ (accessed Jul. 19, 2021).